\documentclass[twocolumn,prl,aps,twocolumn]{revtex4-1}
\usepackage[latin9]{inputenc}
\usepackage{amsmath}
\usepackage{amssymb}
\usepackage{graphicx}
\usepackage{makecell}
\usepackage[unicode=true,pdfusetitle,
 bookmarks=false,
 breaklinks=false,pdfborder={0 0 1},backref=false,colorlinks=false]
 {hyperref}
\hypersetup{
 bookmarksnumbered=false,bookmarksopen=false}

\makeatletter
\@ifundefined{textcolor}{}
{%
 \definecolor{BLACK}{gray}{0}
 \definecolor{WHITE}{gray}{1}
 \definecolor{RED}{rgb}{1,0,0}
 \definecolor{GREEN}{rgb}{0,1,0}
 \definecolor{BLUE}{rgb}{0,0,1}
 \definecolor{CYAN}{cmyk}{1,0,0,0}
 \definecolor{MAGENTA}{cmyk}{0,1,0,0}
 \definecolor{YELLOW}{cmyk}{0,0,1,0}
}

\usepackage{color}

\newcommand{\ehbar}{\hbar_{\mathrm{eff}}}

\@ifundefined{textcolor}{}{%
 \definecolor{BLACK}{gray}{0}
 \definecolor{WHITE}{gray}{1}
 \definecolor{RED}{rgb}{1,0,0}
 \definecolor{GREEN}{rgb}{0,1,0}
 \definecolor{BLUE}{rgb}{0,0,1}
 \definecolor{CYAN}{cmyk}{1,0,0,0}
 \definecolor{MAGENTA}{cmyk}{0,1,0,0}
 \definecolor{YELLOW}{cmyk}{0,0,1,0}
}

\usepackage{txfonts}

\makeatother

\begin{document}
\title{Super-exponential scrambling of Out-of-time-ordered correlators}
\author{Wen-Lei Zhao}
\email{wlzhao@jxust.edu.cn}
\affiliation{School of Science, Jiangxi University of Science and Technology, Ganzhou 341000, China}

\author{Yue Hu}
\affiliation{Guangdong Provincial Key Laboratory of Quantum Engineering and Quantum Materials, GPETR Center for Quantum Precision Measurement, Frontier Research Institute for Physics and SPTE, South China Normal University, Guangzhou 510006, China}

\author{Zhi Li}
\email{lizhi_phys@m.scnu.edu.cn}
\affiliation{Guangdong Provincial Key Laboratory of Quantum Engineering and Quantum Materials, GPETR Center for Quantum Precision Measurement, Frontier Research Institute for Physics and SPTE, South China Normal University, Guangzhou 510006, China}

\author{Qian Wang}
\email{qwang@zjnu.edu.cn}
\affiliation{Department of Physics, Zhejiang Normal University, Jinhua 321004, China}

\begin{abstract}
Out-of-time-ordered correlators (OTOCs) are an effective tool in characterizing black hole chaos, many-body thermalization and quantum dynamics instability. Previous research findings have shown that the OTOCs' exponential growth (EG) marks the limit for quantum systems. However, we report in this letter a periodically-modulated nonlinear Schr\"odinger system, in which we interestingly find a novel way of information scrambling: super-EG. We show that the quantum OTOCs' growth, which stems from the quantum chaotic dynamics, will increase in a super-exponential way. We also find that in the classical limit, the hyper-chaos revealed by a linearly-increasing Lyapunov exponent actually triggers the super-EG of classical OTOCs. The results in this paper break the restraints of EG as the limit for quantum systems, which give us new insight into the nature of information scrambling in various fields of physics from black hole to many-body system.
\end{abstract}
\date{\today}

\maketitle

{\color{blue}\textit{Introduction.---}}
Akin to quantum butterfly effects, quantum scrambling is the process of encoded information spreading from local degrees of freedom to multiple degrees of freedom, hence comes its unattainability by measuring the local operators~\cite{Swingle19,Lewis19}. However, through the time evolution of out-of-time-ordered correlators (OTOCs) ~\cite{Xu20,Prakash20,Sekino08,Landsman19,Rozenbaum20}, this elusive process can be well quantified, which explains the enthusiastic and extensive study of OTOCs in many frontiers of physics such as quantum holography, quantum chaos and black hole physics. Recent research has proved that OTOCs act as an effective indicator of quantum phase transition~\cite{Dag19,Sahu19,Wangqian19,Heyl2018}, many-body localization~\cite{Rammensee18,Ray18,Huang16} and quantum entanglement~\cite{Garttner18}. Besides, experimental progress has made it possible to observe the OTOCs in atomic-optics setups~\cite{Rammensee18,Ray18} and nuclear spins~\cite{Li2017}. Up to now, a wide range of OTOCs with logarithmic, power-law or exponential growth (EG) has been found in various systems including many-body systems and quantum chaotic systems.

As proposed in landmark studies of quantum chaos, the OTOCs can be used to describe the exponential instability in quantum dynamics~\cite{Hashimoto2017,Mata2018B,Fortes19,Dora2017,Rozenbaum17,Rozenbaum19,Yan19}. As for systems with the well-defined classical limit, the EG will occur within Ehrenfest time with a rate determined by the classical Lyapuonv exponent~\cite{Mata2018B,Fortes19,Rozenbaum17,Lakshmi19}. Interestingly, the mathematically verified correlation between OTOCs and Loschmidt echo provides theoretical basis for the irreversibility of scrambling dynamics~\cite{Zurek20}. Note that, the boundary set by chaos dynamics for exponential scrambling of OTOCs, which means that an exponent actually marks the greatest rate of increase for OTOCs in a quantum system, is obtained by the conjecture of thermalization in quantum systems with a large number of degrees of freedom~\cite{Murthy19,Maldacena16}. At present, massive research efforts have been focused on how many-body chaos affects the dynamics of OTOCs. For instance, a recent study has shown that the OTOCs' EG is equal to the classical Lyapunov exponent in the presence of interatomic interaction~\cite{Mata2018}. However, we noticed that the growth rate of OTOCs in previous systems fails to break the exponential limit. Therefore, is it safe to say that the EG actually tops all OTOCs' rates of growth in quantum systems?

{\color{blue}\textit{Model and results.---}}
We consider a Schr\"odinger system with the temporally modulated nonlinear interaction, the corresponding Hamiltonian reads~\cite{Zhao16,Zhao19}
\begin{equation}\label{NSEHmail}
{\rm H}=\frac{p^2}{2}+g|\psi(\theta,t)|^2\sum_j\delta(t-j)\;,
\end{equation}
where the angular momentum operator $p=-i\ehbar \partial/\partial \theta$ with $\ehbar$ being the effective Planck constant, $\theta$ is angle coordinate, $g$ denotes the nonlinear interaction strength. All variables are properly scaled and thus in dimensionless units.
An arbitrary quantum state is expanded in terms of the complete basis of angular momentum operator ($p |\varphi_n\rangle = n\ehbar |\varphi_n\rangle$), i.e., $|\psi\rangle = \sum_n \psi_n |\varphi_n\rangle$, hence, it is periodical in $\theta$, i.e., $\psi(\theta)=\psi(\theta+2\pi)$.
The one-period evolution operator from time $t$ to $t+1$ is given by $U(t,t+1)=\exp\left(-i{p^2}/{2\ehbar}\right)\exp\left[-i{g|\psi(\theta,t)|^2}/{\ehbar}\right]$. The quantum OTOCs are defined as the average of the squared commutator, i.e., $C(t) = -\langle [\hat{A}(t), \hat{B}(0)]^2\rangle$, where $\hat{A}(t)=e^{i Ht}A(0)e^{-i Ht}$ is the time-dependent operator in Heisenberg picture, with $\langle \cdot \rangle$ being the average of the initial state.
In the many-body systems, the average of $C(t)$ comes from thermal states. For the quantum mapping systems, however, there is no definition of thermal states, as the temperature tends to be infinitely large after long time evolution~\cite{Rigol14}. Indeed, our previous investigations have proved that the system in Eq.~\eqref{NSEHmail} exhibits the unbounded heating, which is quantified by the exponentially-fast growth of mean energy $\langle p^2\rangle$~\cite{Zhao16,Zhao19}.

Here, we consider the case of a pure state, i.e., a Gaussian wavepacket $\psi(\theta,0)=(\sigma/\pi)^{1/4}\exp(-\sigma\theta^2/2)$.
As in ref.~\cite{Rozenbaum17}, $\hat{A}=\hat{B}=p$, namely
\begin{equation}\label{QOTOCDef}
C(t) = - \left\langle [{p}(t), {p}(0)]^2 \right\rangle\;.
\end{equation}

\begin{figure}[htbp]
\begin{center}
\includegraphics[width=7.5cm]{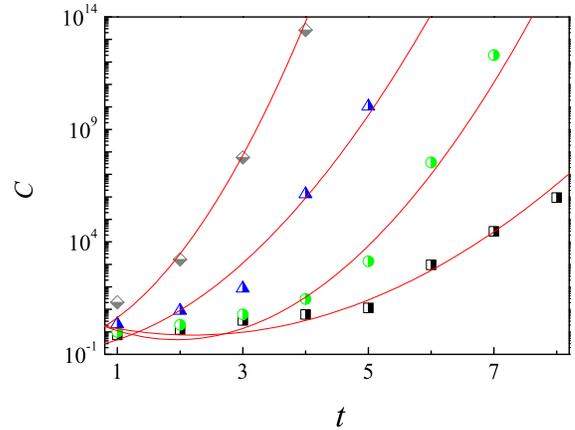}
\caption{(color online). (on log-linear scale) Quantum OTOC $C(t)$ versus time with $\ehbar=0.6$ for $g =1.3$ (squares), $1.5$ (circles), $2.0$ (triangles) and $3.0$ (diamonds). Red solid lines indicate our theoretical prediction in Eq.~\eqref{TheoryOTOC}. The width of Gaussian wavepacket is $\sigma=1.0$. \label{OTOC}}
\end{center}
\end{figure}

Our main result is the analytical prediction of the Super-EG scrambling of the quantum OTOCs
\begin{equation}\label{TheoryOTOC}
C(t) \propto \exp\left[\alpha \gamma_{} t+ \beta \ln\left(\frac{g}{\pi\ehbar}\right)^2t + \eta\gamma_{} t^2 \right] \;,
\end{equation}
where
\begin{equation}\label{GRN}
\gamma_{} = \ln\left[1+ \left(\frac{g \mathcal{N}_0}{\pi \ehbar}\right)^2\right]\;,
\end{equation}
$\alpha$, $\beta$ and $\eta$ are prefactors and $\mathcal{N}_0$ is the normalization constant of the initial state (usually $\mathcal{N}_0=1$). Numerical results of OTOCs are in good agreement with the analytical expression (see Fig.~\ref{OTOC}). It is worth noting that EG is usually believed to be the boundary of the scrambling of OTOCs in chaotic systems~\cite{Maldacena16}, therefore, our finding of the super-EG scrambling sheds new light in the field of quantum information~\cite{Dressel18,Alonso19}. We have also investigated, both numerically and analytically, the OTOCs defined as $C(t)=-\langle[\theta(t),p]^2\rangle$. Interestingly, we found the scaling
\begin{equation}
C(t)=\mu(t)N\exp(\nu\gamma t)\;,\label{OTOCScal}
\end{equation}
where $\mu$ is the time-dependent coefficient, $N$ is number of basis, namely the dimension of the system, $\nu$ is a constant, and the growth rate reads
\begin{equation}\label{GRate}
\gamma= \ln \left\{1+\left[\frac{g\tilde{\mathcal{N}}(t)}{\pi \ehbar}\right]^2\right\}\;
\end{equation}
with $\tilde{\mathcal{N}}(t)=\langle \psi(t)|\theta^2|\psi(t)\rangle$~\cite{Supple}. It is obvious that the $C(t)$ approaches to infinite with the increase of the dimension of the system $N$, which demonstrates that there is no bound on the OTOCs in our system.

The bound of the exponential growth of OTOCs applies to many-body systems with finite temperature where the thermal states are well defined. For such systems, the two-point correlator in OTOCs saturates rapidly. The exponential growth of OTOCs mainly results from the four-point correlator~\cite{Maldacena16}. It has proved that the periodically-driven systems are equivalent to the systems with infinite temperature~\cite{Rigol14}, for which the thermal states cannot be well defined without an effective Hamiltonian. These essential differences lead to the inconsistency between the super-EG of OTOCs in our system and the bound of chaos in ref.~\cite{Maldacena16}. Remarkably, for periodically-driven systems, the two-point correlator contributes mainly to the OTOCs, and the four-point correlator saturates rapidly~\cite{Rozenbaum17,Ueda2018}.
In the following, we will show that the two-point correlator part exhibits the super-exponential growth with time. Here, the super-EG of OTOCs is rooted in the exponentially-fast diffusion of mean energy. It is a clear evidence that the scrambling dynamics is closely related to quantum thermalization, which is highlighted in such interdisciplinary topics as quantum information scrambling and quantum chaos. On the other hand, as a variant of kicked rotor model, our system is an ideal platform for investigating the wavepacet dynamics, such as quantum walk~\cite{Dadras18} and topologically-protected transport~\cite{Ho12} in momentum-space lattice.
Therefore, our finding opens a new prospective in the field of the scrambling dynamics in momentum-space lattice.

We proceed to evaluate the scrambling in the semiclassical limit.
While approaching to the semiclassical limit, the quantum commutator reduces to Possion bracket  $[p(t),p(0)]=\ehbar\{p(t),p(0)\}=\ehbar{\partial p(t)}/{\partial x(0)}$. Then, a natural definition of the classical OTOCs are
\begin{equation}\label{COTOCDef}
C_{cl}(t)=\left\langle \left(\frac{\partial p(t)}{\partial x(0)}\right)^2 \right\rangle\;,
\end{equation}
where $\langle \cdot \rangle$ denotes the average of the ensemble of classical trajectories~\cite{Rozenbaum17}.
In numerical calculations, the classical OTOCs are approximated as $C_{cl}(t)\approx \langle (\delta p (t)/\delta x (0))^2 \rangle $, where $\delta p$ and $\delta x$ denote the difference of the two nearest neighboring trajectories~\cite{Rozenbaum17}.

We have proved that the system in Eq.~\eqref{NSEHmail} is mathematically equivalent to a generalized kicked rotor (GKR) model~\cite{Zhao16,Zhao19,Supple}, whose Hamiltonian takes the form
\begin{align}
\label{nonHamil}
\textrm{H} =\frac{p^2}{2} + \sum_{n=1}^{+\infty} K_n(t)\cos(n\theta)\sum_j(t-j)\;.
\end{align}
The kicking strength dependent on the Fourier components of the quantum state reads
\begin{equation}\label{correlation}
K_n(t) = \frac {g}{\pi} \sum_{m=-\infty}^{+\infty} \psi_m^*(t) \psi_{m+n}(t).
\end{equation}
The classical mapping equation of this GKR model can be formulated as
\begin{equation}\label{stamapping1}
\begin{cases}
p(t+1)-p(t)=  \sum_{n=1}^{+\infty}nK_n(t) \sin\left[ n\theta(t) \right]\\
\theta(t+1)-\theta(t)= p(t+1),
\end{cases}
\end{equation}
where $p(t)$ and $\theta(t)$ indicate the classical momentum and angle variables after the $t$-th kick.

\begin{figure}[htbp]
\begin{center}
\includegraphics[width=8.5cm]{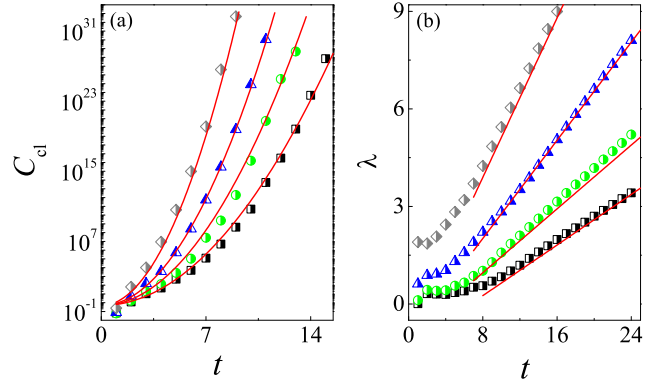}
\caption{(color online). Time dependence of the classical OTOC $C_{cl}(t)$ (a) (on log-linear scale) and the maximal Lyapunov exponent (b) with $g =1.3$ (squares), $1.5$ (circles), $2.0$ (triangles) and $3.0$ (diamonds).
Red dashed lines in (a) and (b) indicate our theoretical predictions in Eqs.~\eqref{COTOC} and~\eqref{CLLyaE}, respectively. Other parameters are the same as in Fig.~\ref{OTOC}.\label{ClaOTOC}}
\end{center}
\end{figure}

Based on the classical mapping equations, we numerically investigate the classical OTOCs for different interaction $g$. In numerical simulations, we set the initial values of $p$ and $\theta$ as random variables with the probability distribution given by Gaussian function in phase space. The difference of initial trajectories is $\delta \theta(0)=10^{-5}$. Interestingly, the classical OTOCs increase in the super-EG way, i.e.,
\begin{equation}\label{COTOC}
C_{cl} (t) \propto \exp \left(\gamma t^2 \right)\;,
\end{equation}
which is in good agreement with our theoretical prediction (see Fig.~\ref{ClaOTOC}(a)).
In addition, we numerically investigate the maximal Lyapunov exponent $\lambda = \lim_{t\rightarrow \infty} \lim_{\delta \theta\rightarrow 0} \langle \log[\delta  p(t)/\delta  \theta(0)]\rangle/t$.
Remarkably, the maximal Lyapunov exponent $\lambda$ linearly increases with time $\lambda (t)\propto \gamma t$ (see Fig.~\ref{ClaOTOC}(a)), which is in consistence with the theoretical prediction in Eq.~\eqref{CLLyaE}. This clearly demonstrates the existence of the hyper-chaotic dynamics~\cite{Andreev19}. Different from previous research, in which the classical OTOCs depend on the $\lambda$ in an exponential way of $C_{cl}(t) \propto e^{\lambda t}$~\cite{Mata2018B,Fortes19,Rozenbaum17,Lakshmi19}, here we get the super-EG with the form $C_{cl}(t)\propto e^{\gamma t^2}$.  Compared with traditional quantum systems~\cite{Guarneri17,Gisin95},  richer physics will be exhibited in the periodically-modulated nonlinear Schr\"odinger system from the perspective of either quantum dynamics or semi-classical dynamics.

{\color{blue}\textit{Theoretical analysis.---}}
It is straightforward to decompose the OTOCs in Eq.~\eqref{QOTOCDef} as
\begin{equation}\label{OTOC0}
C(t)= C_1(t)+C_2(t)-2\text{Re}[C_3(t)]\;,
\end{equation}
where we define the terms in the right of the above equation as
\begin{equation}
\begin{split}
C_1(t)&=  \left\langle\psi(0)\left|{U}^{\dagger}(t) p^{} {U}(t) p^{2} {U}^{\dagger}(t) p \hat{U}(t)\right| \psi(0)\right\rangle\;,\label{OTOC1}\\
C_2(t)&= \left\langle\psi(0)\left|p^{} {U}^{\dagger}(t) p{U}(t){U}^{\dagger}(t) p {U}(t) p\right| \psi(0)\right\rangle\;,\\
C_3(t)&=\left\langle\psi(0)\left| {U}^{\dagger}(t)  p^{}  {U}(t) p^{} {U}^{\dagger}(t)  p \hat{U}(t) p\right| \psi(0)\right\rangle\;.
\end{split}
\end{equation}
Our numerical investigation shows that the contribution of the first term $C_1(t)$  is larger than the other two about several orders of magnitude, which means $C(t)\approx C_1(t)$~\cite{Ueda2018}. The two-point correlation function quantifies the quantum irreversibility measured by the expectation value of $p^2$ under the perturbation of $p$ at the time $t$. Hereinafter, we will show how the exponentially-fast diffusion of mean energy ($\langle p^2\rangle\propto e^{\gamma t}$) induces the super-EG of $C_1(t)$.

The definition of $C_1(t)$ in Eq.~\eqref{OTOC1} means that there are four steps in the calculation of this part. The first step is the forward time evolution of an initial state from $t_0$($=0$) to $t_{+}=t_0 +t$, which yields the quantum state  $|\psi(t_{+})\rangle={U}(t)|\psi(0)\rangle$.
In the second step, the operator $p$ is exerted on the state $|\psi(t_{+})\rangle$, i.e., $|\tilde{\psi}(t_{+})\rangle\equiv p|\psi(t_{+})\rangle= {p}{U}(t)|\psi(0)\rangle$.
The norm of this state has the expression $\mathcal{N}_{t_{+}}=\langle \tilde{\psi}(t_{+})|\tilde{\psi}(t_{+})\rangle=\langle \psi(0)|{U}^{\dagger}(t) {p}^2{U}(t)|\psi(0)\rangle$, which is just the mean energy at the time $t_{+}$. Our previous investigation has shown that the mean energy exponentially increases~\cite{Zhao16,Zhao19}, i.e.,
\begin{equation}\label{normt}
\mathcal{N}_{t_{+}}\propto \mathcal{N}_{0} \exp(\gamma t)=\exp(\gamma t)\;,
\end{equation}
where we adopted the condition $\mathcal{N}_{0}=\langle \psi(0)|\psi(0)\rangle=1$.
The third step is the time reversal from $t_{+}$ to $t_0 $ ($t$ steps), which results in a state $|\varphi(t_{0})\rangle \equiv{U}^\dagger(t)|\tilde{\psi}(t_{+})\rangle={U}^\dagger(t)\hat{p}\hat{U}(t)|\psi(0)\rangle$. Finally, in the fourth step, one can obtain the expectation value of $p^2$ with the state $|\varphi(t_{0})\rangle$, i.e., $C_1(t)\approx\langle \varphi(t_{0})|p^2|\varphi(t_{0})\rangle$ $=\langle \tilde{\psi}(t_{+})|{U}(t)p^2{U}^\dagger(t)|\tilde{\psi}(t_{+})\rangle$.

Note that, the process of time reversal (the third step) can be viewed as a process starting from a new initial state $|\tilde{\psi}(t_+)\rangle$ normalized to $\mathcal{N}_{t_{+}}$, and then evolving into a quantum state $|\varphi(t_{0})\rangle $ in a time interval $t$. The two-point correlator  $C_1(t)$ is just the  ``mean energy'' of the state $|\varphi(t_{0})\rangle $, which follows the exponentially-fast way
\begin{equation}\label{OTOC2}
C_1(t) \propto \mathcal{N}_{t_{+}}\exp(\gamma_{t_{+}} t) \propto \exp(\gamma t+\gamma_{t_{+}} t)\;
\end{equation}
with the rate $\gamma_{t_{+}}=\ln\left[1+\left({g{{\cal{N}}_{t_{+}}}}/{\pi\ehbar}\right)^2\right]$~\cite{Zhao16,Zhao19}.
Taking Eq.~\eqref{normt} into account, the ${\gamma}_{t_{+}}$ has the expression
\begin{align}\label{GRC2}
{\gamma}_{t_{+}}&\propto\ln\left[1+\left(\frac{g }{\pi\ehbar}\right)^2e^{2\gamma t}\right]\\\nonumber
&\propto \ln\left(\frac{g}{\pi\ehbar}\right)^2 + \gamma t\;,
\end{align}
where we use the condition $e^{2\gamma t} \gg 1$.
As a consequence, the OTOCs take the form
\begin{equation}\label{OTOC4}
C(t) \approx C_1(t)\propto\exp\left[\alpha \gamma t+ \beta \ln\left(\frac{g}{\pi\ehbar}\right)^2t + \eta\gamma t^2 \right] \;,
\end{equation}
where the prefactors $\alpha$, $\beta$ and $\eta$
can not be exactly obtained, since in the process of derivation we have used approximations in Eqs.~\eqref{normt},~\eqref{OTOC2} and~\eqref{GRC2}. According to the above analysis, the super-EG of OTOCs is mainly caused by the positive feedback mechanism of the temporally-modulated interaction, which is the key point in this letter. The exponential increase of the kicking strength, which is absent in the traditional kicked rotor model, is responsible for the appearance of super-EG in the classical limits of the system.

Next, we analyze the time evolution of the classical OTOCs. For the kicked rotor model, the maximum Lyapunov depends on the kicking strength by the way of $\lambda \propto \ln(K)$~\cite{Chirikov79,Rozenbaum17}. In the GKR model, the definition of one of the components of the kicking strength $K_n$ in Eq.~\eqref{correlation} indicates that the $K_n$ is the quantum correlation in momentum space. A significant feature of this system is the exponential localization of quantum states, i.e., $|\psi(p)|^2\sim \exp(-|p|/\xi)$ with $\xi$ being the localization length. Through simple calculation, one can obtain that the quantum correlation also has the exponential decay, i.e., $K_n \propto \exp(-|p|/\xi)$~\cite{Zhao16,Zhao19}.
Then, a rough estimation of the kicking strength in the GKR model (see Eq.~\eqref{stamapping1}) is $K(t) = \sum_n n K_n(t)\propto \sum_n n e^{-|n|\ehbar/\xi}\propto \xi$.
Previously, we have predicted the exponentially-fast increase of the localization length $\xi \propto e^{\gamma t}$ by means of the hybrid quantum-classical theory~\cite{Zhao16,Zhao19}. As a consequence, the kick strength exponentially grows with time, i.e., $K \propto e^{\gamma t}$. Accordingly, the Lyapunov exponent linearly increases with time
\begin{equation}\label{CLLyaE}
\lambda \propto \ln(K) \propto \gamma t\;.
\end{equation}
For strong chaotic case, the classical OTOCs exponentially increase with the growth rate proportional to Lyapunov exponent~\cite{Mata2018B,Fortes19,Rozenbaum17,Lakshmi19}, i.e.,
$C_{cl}(t) \propto e^{\lambda t}$.
Taking Eq.~\eqref{CLLyaE} into account, one can get $C_{cl}(t) \propto \exp(\gamma t^2)$,
which is confirmed by our numerical results (see Fig.~\ref{ClaOTOC}).

{\color{blue}\textit{Conclusion and prospects---}}We have proved the existence of OTOCs' super-EG. The results not only give evidence to verify the association between quantum hyper-chaos and quantum scrambling,  but also confirm the super-exponential sensitivity of quantum dynamics to initial conditions. Since the conventional theory has it that a EG boundary is imposed on the chaotic dynamics, our finding, by breaking traditional restraints, bears great significance in the research of information scrambling and quantum chaos~\cite{Ray18}. More importantly, the super-EG is the fastest growth rate of OTOC today, for which, the underlying mechanism is the positive feedback of the periodic modulated interaction. Our finding helps review the issue of the chaotic systems' boundary.

Experimentally, due to the generality of the nonlinear Schr\"{o}diner system, our findings will serve as a universal theory in broad fields including the cold atomic gases~\cite{Dalfovo99}, nonlinear optics~\cite{Szameit10,Longhi11,Blomer06} and complex Ginzburg-Landau equation in condensed-matter physics~\cite{Anderson2002}. More remarkably, the interaction in the above-mentioned systems features high controllability~\cite{Inouye98,Kevrekidis03,Theis04,Anderson2002,Chin10,Szameit10,Longhi11,Blomer06}, which will facilitate the experimental realization and observation of the predictions.

\begin{acknowledgments}
We are grateful to Jie Liu and Jiaozi Wang for valuable suggestions
and discussions. This work was supported by the Natural Science Foundation of China under Grant Nos. 12065009 and 11704132, the PCSIRT
(Grant No. IRT1243), the Natural Science Foundation of Guangdong Province (No. 2018A030313322, and No. 2018A0303130066), the KPST of Guangzhou (Grant No. 201804020055) and science and technology program of Guangzhou (No. 2019050001)
\end{acknowledgments}

\pagebreak
\clearpage
\begin{center}
\textbf{\large Supplemental Material:\\ Super-exponential scrambling of Out-of-time-ordered correlators}
\end{center}
\setcounter{equation}{0} \setcounter{figure}{0} \setcounter{table}{0}
\setcounter{page}{1} \makeatletter \global\long\def\theequation{S\arabic{equation}}
 \global\long\def\thefigure{S\arabic{figure}}
 \global\long\def\bibnumfmt#1{[S#1]}
 \global\long\def\citenumfont#1{S#1}

\section{S1. Details about the mathematical equivalence between the nonlinear Schr\"odinger system and the GKR model}

The periodically modulated nonlinear Schr\"odinger system reads
\begin{equation}\label{NSEHmail_SM}
{\rm H}=\frac{p^2}{2}+g|\psi(\theta,t)|^2\sum_j\delta(t-j)\;.
\end{equation}
For a symmetric initial state ($\psi(\theta,0)=\psi(-\theta,0)$), since the kicking evolution operator $U_K=\exp\left[-i{g|\psi(\theta,t)|^2}/{\ehbar}\right]$ is wavefunction-dependent, the quantum state will preserve the symmetry in the duration of evolution, i.e., $\psi(\theta,t)=\psi(-\theta,t)$. The wavefunction can be expanded as $\psi(\theta,t)=\sum_{n=- \infty}^{+\infty} \psi_n(t) e^{in\theta}/\sqrt{2\pi}$, therefore one can obain
\begin{align}
\label{modsqur}
\begin{split}
|\psi(\theta,t)|^2&= 2\sum_{n=- \infty}^{+\infty} Y_n(t) e^{in\theta}\\
&= 2 Y_0 + 4\sum_{n=1}^{+\infty} Y_n(t) \cos(n\theta),
\end{split}
\end{align}
where
\begin{equation}\label{DefCorrel}
Y_n(t) = \frac 1{4\pi} \sum_{m=-\infty}^{+\infty} \psi_m^*(t) \psi_{m+n}(t).
\end{equation}
The expression Eq.~\eqref{DefCorrel} describes the correlation of the quantum state in the momentum space. Besides, one can get $Y_n(t)=Y_{-n}(t)$~\cite{Zhao16,Zhao19}. By plugging Eq.~\eqref{modsqur} into Eq.~\eqref{NSEHmail_SM}, we have
\begin{align}\label{nonHamil}
\textrm{H} =\frac{p^2}{2} + \sum_{n=1}^{+\infty} K_n(t)\cos(n\theta)\sum_j(t-j)\;.
\end{align}
where the kicking strength $K_n(t)=4g Y_n(t)$. Since the term with $Y_0=1/4\pi$ only contributes a global phase in the evolution, which has no physical effects, we can drop it safely.

\begin{figure*}[htbp]
\begin{center}
\includegraphics[width=14.0cm]{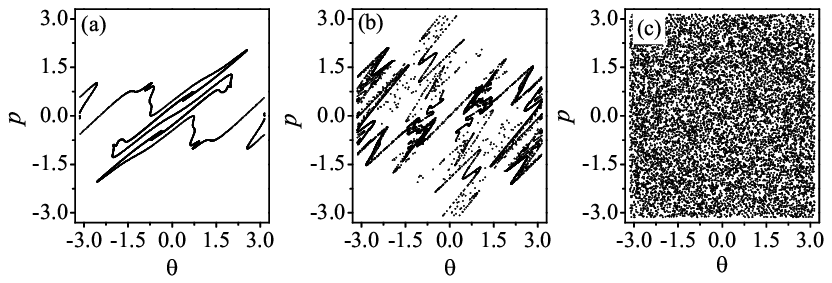}
\caption{Phase space portrait of the classical GKR model for an ensemble of $N=10^4$ trajectories with $t=3$ (a), $5$ (b) and $15$ (c). The parameters are $g=1.5$ and $\ehbar=0.6$.\label{PSpace}}
\end{center}
\end{figure*}

We numerically investigate the phase space of classical trajectories. Our results show that, for a specific value of $g$, the classical phase space exhibits regular diffusion of trajectories for short time evolution (e.g., $t=3$  in Fig.~\ref{PSpace}(a)), the coexistence of both the regular diffusion and the chaotic diffusion for intermediate time interval (e.g., $t=5$  in Fig.~\ref{PSpace}(b)), and the full chaotic diffusion after long enough time evolution (e.g., $t=15$  in Fig.~\ref{PSpace}(c)). This clearly demonstrates the regular-to-chaotic transition of the classical dynamics, which stems from the time-dependent increase of the GKR model's kick strength. The regular behavior leads to the early-time deviation of the Lyapunov exponent from the theoretical prediction in Eq.(18) of the main text, whose validity is guaranteed under the strong chaotic condition (see Fig.1(a) in main text).

\section{S2. Quantum and classical OTOCs for short time evolution}

\begin{figure}[htbp]
\begin{center}
\includegraphics[width=8.0cm]{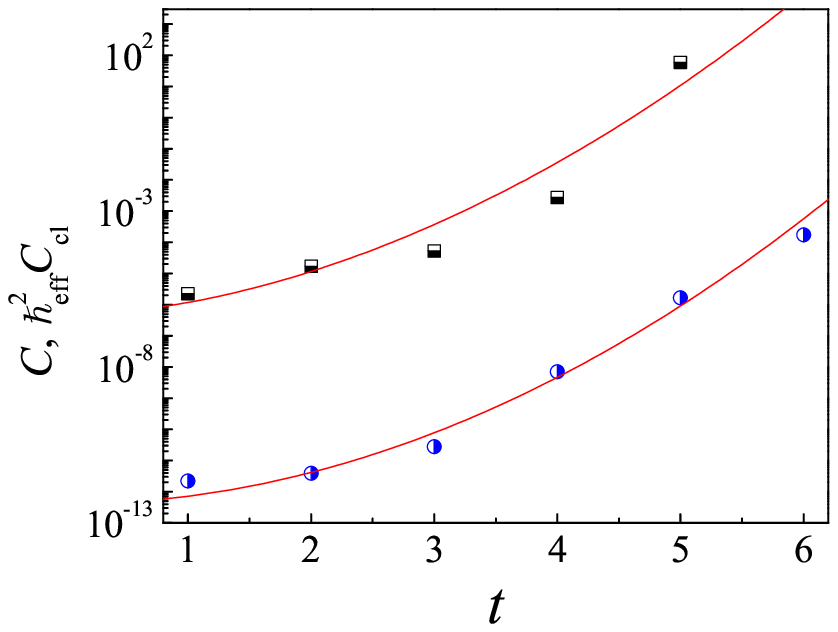}
\caption{(color online). (on log-linear scale) Time dependence of the $C(t)$ (squares) and $\ehbar^2 C_{cl}(t)$ (circles) with $\ehbar=0.001$ and $g =0.01$ . Dashed lines (in red) indicate our theoretical predictions. The width of Gaussian wavepacket is $\sigma=10$.\label{OTOCQC}}
\end{center}
\end{figure}

For systems described by Schr\"odinger equation, the quantum OTOCs are consistent with its classical counterpart, i.e., $C(t)=\ehbar^2 C_{cl}(t)$ in the semiclassical limit ($\ehbar \ll 1$)~\cite{Rozenbaum17,Mata2018}. The comparison between $C(t)$ and $\ehbar^2 C_{cl}(t)$ for our system is shown in Fig.~\ref{OTOCQC}. One can see that, during short time interval, the quantum OTOC is larger than its classical counterpart, which is in sharp contrast to that of the Schr\"odinger system~\cite{Zhao16,Zhao19}. Unfortunately, since the wavepackets spread in the super-exponential way, the long-time evolution faces the severer computation limit on system size. Our theoretical prediction of quantum OTOCs, i.e., $C(t)\propto\exp\left[\alpha \gamma_{} t+ \beta \ln\left(\frac{g}{\pi\ehbar}\right)^2t + \eta\gamma_{} t^2 \right]$ (see Eq.(3) in the main text), yields $C(t)\propto\exp\left(\eta\gamma_{} t^2 \right)$ after long time evolution, which is consistent with the time dependence of classical OTOCs $C_{cl} (t) \propto \exp \left(\gamma t^2 \right)$ (see Eq.(9) in the main text) regardless of the factor $\eta$.

\section{S3. Details about the quantum evolution}
Since the exact form of the temporal modulation function is the periodical-delta kicks, to get the Floquet operator, one can use the standard method of the time integral. The evolution of a quantum state from $t=n$ to $t=n+1$ can be separated into two steps: i). the free evolution from $t=n^+$ to $t=(n+1)^-$, where the superscripts `+' (`-') indicate the time immediately after (before) the  $n$-th kick, i.e., $|\psi[(n+1)^-]\rangle = U_f|\psi(n^+)\rangle $; and ii). the kick evolution during the infinitely small time interval from  $t=(n+1)^-$ to $t=(n+1)^+$, i.e., $|\psi[(n+1)^+]\rangle = U_K|\psi[(n+1)^-]\rangle $. The integral for the free evolution yields the Floquet operator $U_f = \exp\left({-i p^2}/{2\ehbar}\right)$. For the integral of delta kick, it is straightforward to get the Floquet operator $U_K= \exp\left\{{-i g|\psi[\theta,(n+1)^-]|^2}/{\ehbar}\right\}$. Then, one can get the one-period evolution operator
\begin{equation}\label{FlqOper-SM}
U=U_fU_K = \exp\left(\frac{-i p^2}{2\ehbar}\right)
\exp\left[\frac{-i g|\psi( \theta,t)|^2 }{\ehbar}\right]\;.
\end{equation}

\begin{figure*}[htbp]
\begin{center}
\includegraphics[width=16.0cm]{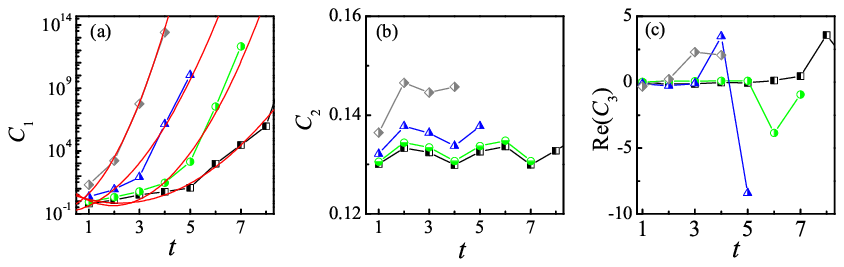}
\caption{(color online). Time dependence of the $C_1$ (a), $C_2$ (b) and $\text{Re}[C_3]$ (c)   with $\ehbar=0.6$ for $g =1.3$ (squares), $1.5$ (circles), $2.0$ (triangles) and $3.0$ (diamonds). Red solid lines indicate our theoretical prediction in Eq.~\eqref{TheoryOTOC}. The width of Gaussian wavepacket is $\sigma=1.0$.\label{OTOC23}}
\end{center}
\end{figure*}

Our main result is the analytical prediction of the Super-EG scrambling of the quantum OTOCs
\begin{equation}\label{TheoryOTOC}
C(t) \propto \exp\left[\alpha \gamma_{} t+ \beta \ln\left(\frac{g}{\pi\ehbar}\right)^2t + \eta\gamma_{} t^2 \right] \;,
\end{equation}
where
\begin{equation}\label{GRN}
\gamma_{} = \ln\left[1+ \left(\frac{g \mathcal{N}_0}{\pi \ehbar}\right)^2\right]\;,
\end{equation}
$\alpha$, $\beta$ and $\eta$ are prefactors and $\mathcal{N}_0$ is the normalization constant of the initial state (usually $\mathcal{N}_0=1$). We cannot analytically get the values of  $\alpha$,  $\beta$, $\eta$ and the proportionality constant. To fit the numerical results in Fig.~\ref{OTOC23}(a), we select a set of $(\alpha,\beta,\eta)$ and a set of proportionality constants, with the aim of making the curve of the analytical formula in Eq.~\eqref{TheoryOTOC} and the numerical results a good match. Same method is taken for the red curves in Fig.2 in the main text. Therefore, the values of $\alpha$, $\beta$, $\eta$ and the proportionality constant are dependent on the value of $g$.

\section{S4. Scaling of the OTOCs $C(t)=-\langle [\theta(t),p]^2\rangle$}
\begin{figure}[htbp]
\begin{center}
\includegraphics[width=8.0cm]{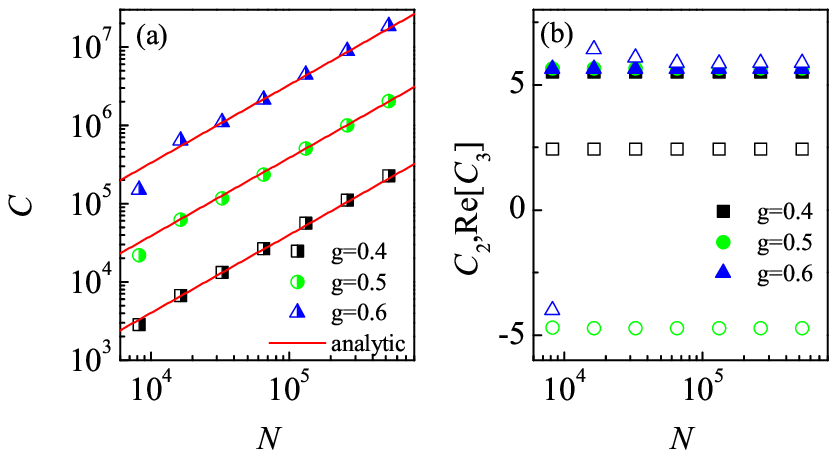}
\caption{(color online). (a) Dependence of $C_1(t)$ (a), $C_2(t)$ (b), and $\textrm{Re}[C_3(t)]$ (b) on $N$  with $t=7$ for $g=0.4$ (squares), $0.5$ (circles) and $0.6$ (triangles). In (a): red lines denote our theoretical prediction in Eq.~\eqref{OTOCScal}. In (b): solid (empty) symbols denote $C_2(t)$ ($\textrm{Re}[C_3(t)]$). The parameter is $\ehbar=0.6$.
 The width of Gaussian wavepacket is $\sigma=10$.\label{Scaling}}
\end{center}
\end{figure}

In this section, we show the scaling of the OTOCs $C(t)=-\langle [\theta(t),p]^2\rangle$ with the dimension of the system.  Theoretically, the law of time-dependent takes the form
\begin{equation}
C(t)=\mu(t)N\exp(\nu \gamma t)\;,\label{OTOCScal}
\end{equation}
where $\mu$ is a time-dependent coefficient, $N$ is number of basis, $\nu$ is a constant, and the growth rate $\gamma$  has the expression
\begin{equation}\label{GRate-SM}
\gamma= \ln \left\{1+\left[\frac{g\tilde{\mathcal{N}}(t)}{\pi \ehbar}\right]^2\right\}\;
\end{equation}
with $\tilde{\mathcal{N}}(t)=\langle \psi(t)|\theta^2|\psi(t)\rangle$. Equation~\eqref{OTOCScal} means that, at a specific time $t$, the $C(t)$ scales linearly with the dimension of the system, hence it approaches to infinity with the increase of $N$. The divergence of $C(t)$  demonstrates that there is indeed no bound on the growth of OTOCs in our system. To verify our theoretical prediction, we numerically investigate the  $C(t)$  for different $g$ and $N$. In numerical simulations, the initial state is a Gaussian wavepacket   $\psi(\theta,0)=(\sigma/\pi)^{1/4}\exp(-\sigma\theta^2/2)$. Figure~\eqref{Scaling}(a) shows that, at a specific time $t$, the $C(t)$ linearly increase with $N$, which is in good agreement with our theoretical prediction in Eq.~\eqref{OTOCScal}.

We proceed to show the details in the numerical calculation of the OTOCs with the general definition $C(t)=-\langle [A(t),B]^2 \rangle$. In our present work, we have considered two cases with $(A=p,B=p)$ and $(A=\theta,B=p)$. Numerical procedures of calculating these two different OTOCs are the same.
As shown in the main text, the OTOCs can be decomposed as
\begin{equation}\label{OTOC0-SM}
C(t)= C_1(t)+C_2(t)-2\text{Re}[C_3(t)]\;,
\end{equation}
where we define the terms in the right of the above equation as
\begin{align}
C_1(t)&=  \left\langle\psi_R(0)\left| B^{2} \right| \psi_R(0)\right\rangle\;,\label{OTOC1-SM}\\
C_2(t)&= \left\langle\varphi_R(0)|\varphi_R(0)\right\rangle\;,\label{OTOC2-SM}\\
C_3(t)&=\left\langle\psi_R(0)| B | \varphi_R(0)\right\rangle\;,\label{OTOC3-SM}
\end{align}
with $|\psi_R(0)\rangle={U}^{\dagger}(t) A \hat{U}(t)| \psi(0)\rangle$ and $|\varphi_R(0)\rangle={U}^{\dagger}(t) A {U}(t) B| \psi(0)\rangle$. There are five steps to calculate both $C_1(t)$ and $C_2(t)$ at a specific time $t=t^*$:
\begin{itemize}
\item[(i)] select an initial state $|\psi(0)\rangle$,
\item[(ii)] forward evolution $0\rightarrow t^*$ yields $|\psi(t^*)\rangle = U(t^*,0)|\psi(0)\rangle$,
\item[(iii)] at the time $t^*$, exert the operator $A$ on the state $|\psi(t^*)\rangle$ and get a new state $|\tilde{\psi}(t^*)\rangle=A|\psi(t^*)\rangle$,
\item[(iv)] backward evolution $t^*\rightarrow 0$ yields $|\psi_R(0)\rangle=U^{\dagger}(t^*,0)|\tilde{\psi}(t^*)\rangle$,
\item[(v)] calculate the expectation value $C_1(t^*)=  \left\langle\psi_R(0)\left| B^{2} \right| \psi_R(0)\right\rangle$.
\end{itemize}

To get $C_2(t^*)$, at the step (i), one should exert the operator $B$ on the state $|\psi(0)\rangle$, i.e., $|\varphi(0)\rangle=B|\psi(0)\rangle$. The steps (ii)-(iv) are similar to that of calculating $C_1(t^*)$. At the step (v), one can obtain the term $C_2(t^*)$ (see Eq.~\eqref{OTOC2-SM}) by calculating the inner product of the state $|\varphi_R(0)\rangle$.

At the end of the time reversal, one can use the two state $|\psi_R(0)\rangle$ and $|\varphi_R(0)\rangle$ to get the third term of the OTOCs, i.e., $C_3(t^*)=\left\langle\psi_R(0)|B|\varphi_R(0)\right\rangle$ (see Eq.~\eqref{OTOC3-SM}). The schematic diagram for the evolution progress of the quantum state to calculate the $C_1(t^*)$ with $(A=\theta,B=p)$ is shown in Table~\ref{SchemDigm}.
\begin{table*}[htbp]
\begin{center}
\begin{tabular}{|p{3cm}<{\centering}|c|c|c|c|c|c|c|}
\hline
 & \multicolumn{3}{|c|}{Forward: $t^*$ steps } &$\theta$ action &\multicolumn{3}{|c|}{Backward: $t^*$ steps }\\
\cline{2-8}
\rule{0pt}{12pt} &
\multicolumn{3}{|l|}{ $|\psi(0)\rangle $ $\rightarrow$ $|\psi(1)\rangle$ $\rightarrow$ $\cdots$ $|\psi(t^*)\rangle$} &$|\tilde{\psi}(t^*)\rangle= \theta |\psi(t^*)\rangle$ &
\multicolumn{3}{|l|}{$|\tilde{\psi}(t^*)\rangle$ $\cdots$ $\rightarrow$ $|\psi_R(1)\rangle$ $\rightarrow$ $|\psi_R(0)\rangle$}\\
\hline
\rule{0pt}{12pt}$E(t)=\langle\psi(t)|p^2|\psi(t)\rangle$ & \multicolumn{3}{|l|}{\hspace*{0mm}$E(0) $\hspace*{3mm}$\rightarrow$\hspace*{2mm}$E(1)$ \hspace*{0mm} $\rightarrow$ \hspace*{0mm}$\cdots$\hspace*{2mm}$E(t^*)$} &$\tilde{E}(t^*)=\langle \tilde{\psi}(t^*)|p^2|\tilde{\psi}(t^*)\rangle$ &\multicolumn{3}{|l|}{\hspace*{1mm}$\tilde{E}(t^*)$\hspace*{2mm}$\cdots$ \hspace*{-1mm} $\rightarrow$\hspace*{0.8mm} $E_R(1)$\hspace*{3mm} $\rightarrow$\hspace*{1mm}$E_R(0)$}\\
\hline
\rule{0pt}{12pt}$\mathcal{N}(t)=\langle\psi(t)|\psi(t)\rangle$ & \multicolumn{3}{|l|}{\hspace*{0mm}$\mathcal{N}(0) $\hspace*{3mm}$=$\hspace*{2mm}$\mathcal{N}(1)$ \hspace*{0mm} $=$ \hspace*{2mm}$\cdots$\hspace*{1mm}$\mathcal{N}(t^*)=1$} &$\tilde{\mathcal{N}}(t^*)=\langle \psi(t^*)|\theta^2|\psi(t^*)\rangle$ &\multicolumn{3}{|l|}{\hspace*{1mm}$\tilde{\mathcal{N}}(t^*)$$\cdots$ \hspace*{0.1mm}  $=$\hspace*{3mm}${\cal{N}}_R(1)$\hspace*{3mm} $=$\hspace*{2mm}${\cal{N}}_R(0)$}\\
\hline
\end{tabular}
\caption{Schematic diagram of time evolution to calculate the term $C_1(t^*)=\langle\psi_R(0)|p^2|\psi_R(0)\rangle$. }\label{SchemDigm}
\end{center}
\end{table*}

\begin{figure}[htbp]
\begin{center}
\includegraphics[width=8.5cm]{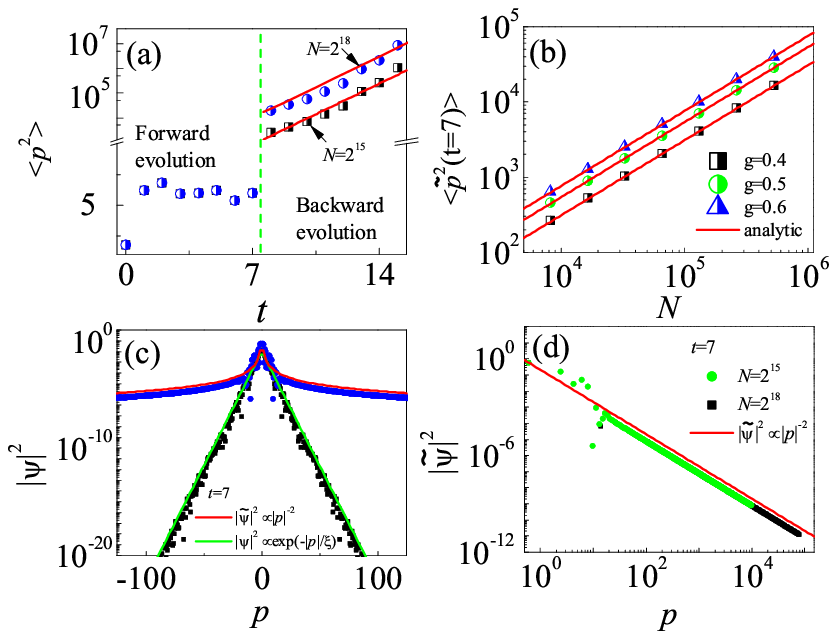}
\caption{(color online). (a) Time dependence of $\langle p^2\rangle$ with $t=7$ and $g=0.6$ for $N=2^{15}$   (squares) and $2^{18}$ (circles). Red solid lines are the theoretical prediction in Eq.~\eqref{EXPDiffTRes}. Green line is the auxiliary line. (b) The $\langle \tilde{p}^2(t=7)\rangle$ versus $N$ with $g=0.4$   (squares), $0.5$ (circles) and $0.6$ (triangles). Red solid lines are the theoretical prediction in Eq.~\eqref{MEngscal}. (c) Momentum distribution at the time $t=7$. Squares (in black) and circles (in blue) indicate the state $|\psi\rangle$ and $|\tilde{\psi}\rangle=\theta|\psi\rangle$, respectively. Red and green lines denote the power-law decay $|\tilde{\psi}(p)|^2\propto |p|^{-2}$ and exponential decay $|\psi(p)|^2\propto \exp(-|p|/\xi)$ with $\xi\approx 2.0$. (d) Momentum distribution for the state $|\tilde{\psi}\rangle$  with $N=2^{15}$ (circles) and $2^{18}$ (squares). We only plot the $|\tilde{\psi}(p)|^2$ of positive $p$, since it is an even function.
Red solid line indicates the power-law decay $\tilde{\psi}(p)\propto |p|^{-2}$. Other parameters are the same as in Fig.~\ref{Scaling}. \label{TEvoltion}}
\end{center}
\end{figure}

Numerically, we find that both $C_2(t)$ and $\textrm{Re}[C_3(t)]$  are negligibly small compared with  $C(t)$ (see Fig.~\ref{Scaling}), which demonstrates that $C(t)\approx C_1(t)$. Then, we proceed to theoretically evaluate the time evolution of $C_1(t)$. Equation.~\eqref{OTOC1-SM} demonstrates that the $C_1(t)$ is just the expectation value of the square of momentum for the state at the end of time reversal, i.e., $C_1(t)=  \left\langle\psi_R(0)\left| p^{2} \right| \psi_R(0)\right\rangle$. We numerically investigate the evolution of the mean square of momentum for a specific time interval $t=t^*$. From Fig.~\ref{TEvoltion}(a), one can see that the values of $\langle p^2\rangle$ remain almost the same during the forward evolution, i.e., $t\leq t^*$. Moreover, the time evolution of the mean energy is independent on the number of basis, which demonstrates the convergence of numerical results. After the action of the $\theta$  operator, i.e., $|\tilde{\psi}(t^*)\rangle=\theta|\psi(t^*)\rangle$, however, the value of $\langle p^2\rangle$  has a clear jump, namely $\langle \tilde{p}^2(t^*)\rangle\gg\langle p^2(t^*)\rangle$. Interestingly, during the time reversal $t>t^*$, the mean energy exponentially increases and has shown clear distinction for different $N$ (see Fig.~\ref{TEvoltion}(a)). Our theoretical prediction of the exponential increase of mean energy is
\begin{equation}\label{EXPDiffTRes}
\langle p^2(t')\rangle = \langle \tilde{p}^2\rangle \exp(\nu \gamma t')\;,
\end{equation}
where $\langle \tilde{p}^2\rangle$  is the mean energy at the time $t=t^*$, $\nu$ is a constant, and the expression of the growth rate is in Eq.~\eqref{GRate-SM}. The detailed derivations of Eq.~\eqref{EXPDiffTRes} are provided in the following. Before that, we will first show the details for the derivation of $C(t)$'s scaling in Eq.~\eqref{OTOCScal}.

Note that, the time reversal starts from the time $t=t^*$ (see Table.~\ref{SchemDigm} for calculating $C_1(t^*)$), therefore $t'=0$  means the time $t=t^*$. Since the time reversal lasts $t^*$ steps (see Table.~\ref{SchemDigm}), the maximum value of $t'$ equals to $t^*$. Accordingly, the mean energy at the end of time reversal is
\begin{equation}\label{MEnergyTResal}
\langle p^2(t=0)\rangle_R = \langle \tilde{p}^2(t^*)\rangle\exp(\nu\gamma t^*)\;,
\end{equation}
which is just the $C_1(t)$  at the time $t=t^*$ (see Eq.~\eqref{OTOC1-SM}). Then, we get the time-dependence of the OTOCs,
\begin{equation}\label{OTOCScal-SM}
C(t)\approx \langle\tilde{p}^2(t)\rangle\exp(\nu\gamma t)\;.
\end{equation}
The scaling of $C(t)$  with $N$  results from the dependence of the mean energy $\langle \tilde{p}^2\rangle$ on $N$. Interestingly, we find that, at a specific time, e.g., $t=t^*$, the $\langle \tilde{p}^2\rangle$ increases with $N$ by the way of
\begin{equation}\label{MEngscal}
\langle \tilde{p}^2(t^*)\rangle = \mu(t^*)N,
\end{equation}
where $\mu$ is a coefficient (see Fig.~\ref{TEvoltion}(b) for $t^*=7$). To reveal the mechanism of such linear scaling, we numerically investigate the distribution of wavepacket at the time $t=t^*$ in the momentum space. Our results show that, after the action of $\theta$ operator on a quantum state, i.e., $|\tilde{\psi}\rangle=\theta|\psi\rangle$, the momentum distribution exhibits the power-law decay (see Figs.~\ref{TEvoltion}(c) and (d))
\begin{equation}\label{PLDecay}
|\tilde{\psi}(p)|^2\propto |p|^{-2}\;,
\end{equation}
for which the mean energy is $\langle \tilde{p}^2\rangle=\int_0^{p_N}p^2|\tilde{\psi}(p)|^{-2}dp \propto p_N$. Then, a rigorous expression of the mean energy is $\langle \tilde{p}^2\rangle=\mu N$  since $p_N=N\ehbar$, which is confirmed by our numerical results in Fig.~\ref{TEvoltion}(b).

The following can explain why the power-law decayed wavefuction appears. In the momentum space, the quantum state $|\tilde{\psi}\rangle$ is expressed as
\begin{equation}\label{MMState}
\langle n|\tilde{\psi}\rangle = \sum_{m}\langle n|\theta|m\rangle \langle m |\psi\rangle\;,
\end{equation}
where the matrix element of $\langle n|\theta|m\rangle$ takes the form
\begin{equation}\label{TetMatrix}
\langle n|\theta|m\rangle=
\begin{cases}
\pi & \text{for}\quad m =n\;,\\
\frac{1}{i(m-n)}&\text{for}\quad m \neq n.
\end{cases}
\end{equation}
The power-law decay of $\langle n|\theta|m\rangle$ is a kind of long-range interaction, which effectively leads to the transition among momentum sites. Even if the quantum state $|\psi\rangle$ is exponentially-localized in the momentum space $|\psi(p)|^2\propto \exp(-|p|/\xi)$ (see Fig.~\ref{TEvoltion}(c)), the action of $\theta$ operator can induce the power-law decay of $|\tilde{\psi}(p)|^2$.

We further show the details for the derivation of the exponential growth of the mean energy in Eq.~\eqref{EXPDiffTRes} during the time reversal. Our previous works have it that, for strong enough nonlinear interaction, the mean energy of our system obeys the iterative equation
\begin{equation}\label{ITEquation}
\langle p^2(t'+1)\rangle \approx \langle p^2(t')\rangle \left[1+ g^2(t')\right]\;,
\end{equation}
where  $g(t')=g|\psi(t')|^2$. Note that, during the time reversal, the norm of the quantum state is a constant, i.e.,  $\langle \psi_R(t')|\psi_R(t')\rangle=\langle \tilde{\psi}(t^*)|\tilde{\psi}(t^*)\rangle=\tilde{\mathcal{N}}(t^*)$.
Thus, a rough estimation of the nonlinear interaction strength is  $g(t')\simeq g\tilde{\mathcal{N}}(t^*)$ with $\tilde{\mathcal{N}}(t^*)=\langle \psi(t^*)|\theta^2|\psi(t^*)\rangle$ (see Table.~\ref{SchemDigm} for $\tilde{\mathcal{N}}(t^*)$). From Eq.~\eqref{ITEquation}, it is straightforward to get the law of the time-dependence of the mean energy
\begin{equation}\label{EXPGEnergy}
\langle p^2(t')\rangle =\langle p^2(t'=0)\rangle\exp(\nu\gamma t')=\langle \tilde{p}^2\rangle \exp(\nu\gamma t')\;
\end{equation}
with
\begin{equation}\label{GRate-SM2}
\gamma= \ln \left\{1+\left[\frac{g\tilde{\mathcal{N}}(t^*)}{\pi \ehbar}\right]^2\right\}\;,
\end{equation}
which is confirmed by our numerical results in Fig.~\ref{TEvoltion}(a).

\end{document}